\documentclass[twoside]{ilcws07}
\usepackage[latin1]{inputenc}
\usepackage[dvips]{graphicx,epsfig,color}
\usepackage{wrapfig,rotating}
\usepackage{amssymb,amsmath,array}

\begin{document}
\title{
%%%%   Paper title goes here  %%%%%%%%%%%%%%
UPDATE ON ION STUDIES{\footnote{This work is supported by the
Commission of the European Communities under the 6th Framework
Programme "Structuring the European Research Area", contract number
RIDS-011899.}}} %%
%***********************************************************************
% AUTHORS INFORMATION AREA
%***********************************************************************
\author{Guoxing Xia and Eckhard Elsen
% Optional short acknowledgment: remove next line if non-needed
%\thanks{This work is supported by the Commission of the European Communities
%under the 6th Framework Programme "Structuring the European Research
%Area", contract number RIDS-011899.}
% DO NOT MODIFY THE FOLLOWING '\vspace' ARGUMENT
\vspace{.3cm}\\
% Addresses and institutions (remove "1- " in case of a single institution)
 DESY, Hamburg, Germany  \\
%% Remove the next three lines in case of a single institution
}
%%***********************************************************************
% END OF AUTHORS INFORMATION AREA
%***********************************************************************
\maketitle
\begin{abstract}
  The effect of ions has received one of the highest priorities in
  R\&D for the damping rings of the International Linear Collider
(ILC). It is detrimental to the performance of the electron damping
ring. In this note, an update concerning the ion studies for the ILC
damping ring is given. We investigate the gap role and irregular
fill pattern in the ring. The ion density reduction in different
fills is calculated analytically. Simulation results are presented.
\end{abstract}
\section{Introduction}
  Ions are recognized as a potential current limitation in storage
rings with negatively charged particle beams \cite{url}. The ions
mainly come from beam-gas collisions. In some circumstances, they
are trapped in the potential well of the beam. They couple to the
motion of the beam and lead to adverse effects such as beam
emittance growth, betatron tune shift and spread, collective
instabilities and beam lifetime reductions \cite{Byrd}.

   There are two kinds of ion effects in electron storage rings.
One is the conventional ion trapping which occurs when the
circulating beam traps ions after multiple turns. It can be cured by
introducing a few successive empty RF bucket (gaps), which are long
compared to the inter-bunch spacing. In this case, the ions are
strongly focused by the passing electron bunches in the beginning
and then over focused in the gap. With a sufficiently large gap, the
ions can be driven to large amplitudes, where they form a diffuse
halo and do not affect the beam. However, in high current storage
rings or linacs with long bunch trains, the ion accumulation during
the passage of a single bunch train may cause a transient
instability which is called fast ion instability (FII) \cite{Tor,
Stupakov}. For the electron damping ring of the ILC, the bunch
intensity is large and the bunch spacing is small and the fast ion
instability is potentially striking \cite{Wolski}. Since the
vertical beam emittance is much smaller (2 pm) than the horizontal
one (0.5 nm), the FII is much more serious in the vertical plane.

  In this note, the linear theory of ion effect is briefly recalled in
section 2. In section 3, the gap effect in the fill is studied and
the ion density reduction due to mini-trains in different fill
patterns is investigated analytically. Section 4 shows the
simulation results of FII for mini-trains. A short summary is given
in the end.
%%{\bfseries We accept submissions in \LaTeX\ only.}
\section{Linear theory of ion effects}
Without gaps in the fill, the ions with a relative molecular mass
greater than \textsl{A} will be trapped in the beam potential,
where,
\begin{equation}
A=\frac{N_{0}r_{p}L_{sep}}{2\sigma_y(\sigma_x+\sigma_y)}
\end{equation}
here, $N_{0}$ denotes the number of particles per bunch, $r_{p}$ the
classical radius of proton, $L_{sep}$  the bunch spacing,
$\sigma_{x,y}$  the horizontal and vertical beam size, respectively.
It can be seen that the minimum trapped mass is closely related to
the beam size. By using the beam parameters of three typical fill
patterns in the ILC damping ring, from case \textsl{A} to case
\textsl{C} as shown in Table 1 \cite{Kuriki}, the minimum trapped
mass along the ring for two fill pattern case \textsl{A} and
\textsl{C} is shown in Figure 1 and Figure 2 respectively. The
number of bunches decreases from \textsl{A} to \textsl{C} while the
particles per bunch increase to maintain a comparable overall
charge. Here we take one sextant of the ring as an example.

% insert table 1 here.
%\begin{wraptable}{l}{0.7\columnwidth}
\begin{table}
\centerline{
\begin{tabular}{|l|c|c|c|}
  \hline
  Beam parameters      & \multicolumn{3}{|c|} {Fill pattern} \\
  \cline{2-4}
          & Case \textsl{A} &  Case \textsl{B} & Case \textsl{C} \\
          \hline
  Number of bunches & 5782 & 4346 & 2767 \\
  Particles per bunch [10$^{10}$] & 0.97 & 1.29 & 2.02\\
  Bunch spacing [bucket] & 2 & 2 & 4 \\
  Number of trains & 118 & 82 & 61 \\
  Bunches per train f$_{2}$ & 0 & 0 & 23 \\
  Gaps between trains g$_{2}$ & 0 & 0 & 28 \\
  Bunches per train f$_{1}$ & 49 & 53 & 22 \\
  Gaps between trains g$_{1}$ & 25 & 71 & 28 \\
  FII char. growth time at train end [10$^{-9}$ s] & 3.922 & 4.527 & 4.030 \\
  FII expo. growth time with 30\% ion freq. spread [10$^{-6}$ s] &
  6.889 & 6.892 & 6.913 \\
  Coherent tune shift at train end & 0.325 & 0.325 & 0.324\\
  \hline
  \end{tabular}
}
 \caption{Typical fill patterns in the ILC damping ring.}
\end{table}

\begin{figure}[htpb]
\begin{minipage}{0.5\columnwidth}
\centerline{\includegraphics[width= 0.9\columnwidth ]{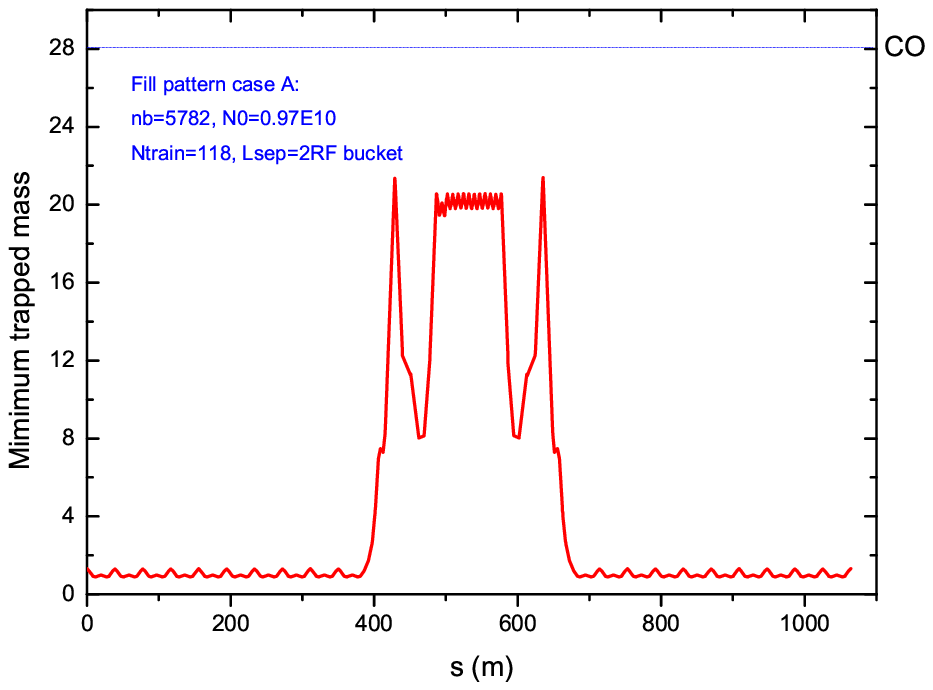}}
\caption{Minimum trapped mass for fill pattern case \textsl{A}.}
\label{Figure:1}
\end{minipage}
\hspace*{0.2cm}
\begin{minipage}{0.5\columnwidth}
\centerline{\includegraphics[width= 0.9\columnwidth ]{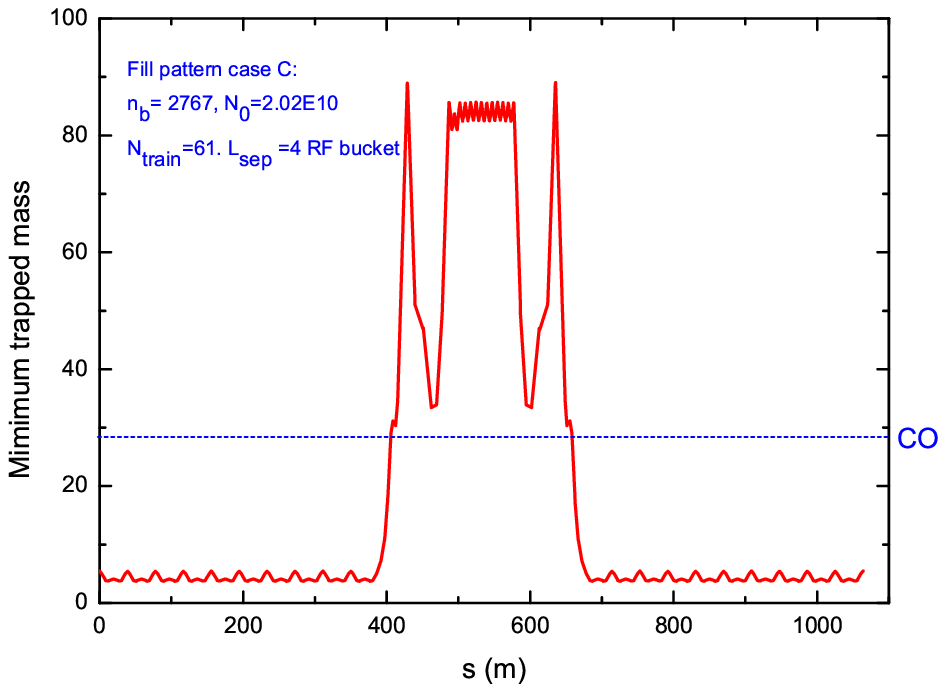}}
\caption{Minimum trapped mass for fill pattern case \textsl{C}.}
\label{Figure:2}
\end{minipage}
\end{figure}
% end of insertion from Shuangshi

It can be seen that for the fill pattern case \textsl{A}, all the CO ions will be
trapped in the beam along the ring. While for the fill pattern case
\textsl{C}, the CO ions can not be trapped in some parts of the
ring.

  The linear theory \cite{Tor, Stupakov} gives characteristic growth
rate of FII which strongly depends on the bunch intensity, number of
bunches, transverse beam size and the residual gas pressure. It can
be estimated as
\begin{equation}
\tau^{-1}_{c}(s^{-1})=5p[Torr]\frac{N^{3/2}_{0}n^{2}_{b}r_{e}r^{1/2}_{p}L^{1/2}_{sep}c}{\gamma\sigma^{3/2}_{y}(\sigma_{x}+\sigma_{y})^{3/2}A^{1/2}\omega_{\beta}}
\end{equation}
where $p$ is the residual gas pressure, $n_{b}$ is the bunch number,
$r_{e}$ and $r_{p}$ are the classical radius of electron and proton
respectively,  $c$ is the speed of light, $\gamma$ is the
relativistic gamma factor, $A$ is the atomic mass number of the
residual gas molecules and $\omega_{\beta}$ is the vertical betatron
frequency. The ion coherent oscillation frequency $\omega_{i}$ is
given by
\begin{equation}
\omega_{i}=(\frac{4N_{0}r_{p}c^{2}}{3AL_{sep}\sigma_{y}(\sigma_{x}+\sigma_{y})})^{1/2}
\end{equation}

However, the ion motion becomes decoherent because the vertical ion
frequency depends on the horizontal position. Furthermore, the
existence of various ion species and the variation of the beam size
along the ring also introduce a spread in the ion oscillation
frequency. Taking into account the ion coherent frequency spread,
the linear theory gives the coupled bunch motion in the bunch train
rising as $y\sim exp(t/\tau_{e})$, and in this case the exponential
growth rate is given by
\begin{equation}
\tau^{-1}_{e}[s^{-1}]=\frac{1}{\tau_{c}}\frac{c}{2\sqrt{2}l_{train}(\Delta\omega_{i})_{rms}}
\end{equation}
where $(\Delta\omega_{i})_{rms}$ denotes the rms spread of the ion
coherent frequency as the function of the azimuthal position around
the ring. $l_{train}=n_{b}L_{sep}$ the bunch train length. For the
baseline design of the ILC damping ring, simulation shows the spread
of ion coherent frequency to be about 30\% \cite{Lanfa}.

If the ions are trapped in the beam potential, they give rise to
additional focusing to the beam. The ion induced coherent tune shift
is given by
\begin{equation}
\Delta{\emph{Q}_{y,coh}}=\frac{\beta_{y}r_{e}\lambda_{ion}C}{\gamma4\pi\sigma_{y}(\sigma_{x}+\sigma_{y})}
\end{equation}
here \textsl{C} is the circumference of ring, $\beta_{y}$ is the
vertical beta function,
$\lambda_{ion}=\sigma_{i}\emph{N}_{0}n_{b}p/k\emph{T}$ is the ion
line density, $\sigma_{i}$ is the ionization cross section (1.86
Mbarn and 0.31 Mbarn for carbon monoxide and hydrogen ions,
respectively at beam energy of 5 GeV). \emph{k} is Boltzmann
constant and \emph{T} is the temperature. By using beam parameters
of the ILC baseline damping ring \cite{Aarons}, the FII
characteristic growth time, exponential growth time with 30\% ion
coherent frequency spread and ion induced coherent tune shift at
bunch train end for a single long bunch train case are analytically
estimated in Table 1. A CO partial pressure of 1 nTorr is assumed
here. It can be seen that the growth time is extremely fast for the
case of one long train. Even with 30\% ion frequency spread, the FII
growth time is still faster than one revolution period $(
22\mu{s})$. The ion induced tune shift is large at a gas pressure of
1 nTorr for CO, so a lower vacuum gas pressure is critical to
alleviate FII effect.
\section{Gap effect in the fill}
  In the previous section, one long bunch train has been assumed and
the ions are trapped by the bunch train. The trapping condition is
disturbed when the fill pattern consists of a number of short bunch
trains (mini-trains) with gaps in between. In the following, we will
analyze the gap effect and ion density in different fill patterns.

  The ions inside the beam are defined as those ions within
$\sqrt{3}\sigma$ of the beam centroid. Note that the growth rate of
FII is proportional to the ion density \cite{Wang}. The diffusion of
the ions during the gaps increases the size of ion cloud and
therefore reduces the ion density. With a gap introduced in the
bunch train, one can estimate the density of the ions in the beam
after the clearing gap as \cite{Byrd}
\begin{equation}
\rho_{i}\approx\frac{\rho_{i0}}{\sqrt{(1+L^{2}_{gap}\omega^{2}_{x})(1+L^{2}_{gap}\omega^{2}_{y})}}
\end{equation}
where $ \rho_{i0}$ is the ion density at the end of the one bunch
train, $L_{gap}$ is the gap length between two adjacent bunch
trains, $\omega_{x,y}$ are the ion oscillation frequencies as
follows
\begin{equation}
\omega^{2}_{x,y}=\frac{2N_{0}r_{p}}{L_{sep}A\sigma_{x,y}(\sigma_{x}+\sigma_{y})}
\end{equation}

For the ILC damping ring, the harmonic number \textsl{h} = 14516,
the circumference \textsl{C} = 6695.057 m. We made an analytic
estimation of the relative ion density reduction versus train gap
spacing. The result is shown in Figure 3. It can be seen that the
relative ion density diminishes with respect to the bunch train gap
spacing. If the gap length is larger than 30 RF bucket, the ion
density is about 10\% of the initial ion density. Beyond 30 RF
bucket, the ion density no longer changes significantly. Taking into
account the transient beam loading effect, train gap should not be
too long. For current ILC damping ring fill patterns, the length of
train gap varies from 25 to 71 RF bucket. Meanwhile, in order to
evaluate the effect of the gaps, an Ion-density Reduction Factor
(\textsl{IRF}) is defined as \cite{Wang}
\begin{equation}
\emph{IRF}=\frac{1}{N_{train}}{\frac{1}{1-exp(-{\tau}_{gap}/{\tau}_{ion})}}
\end{equation}

\begin{figure}[htpb]
\begin{minipage}{0.5\columnwidth}
\centerline{\includegraphics[width= 0.9\columnwidth ]{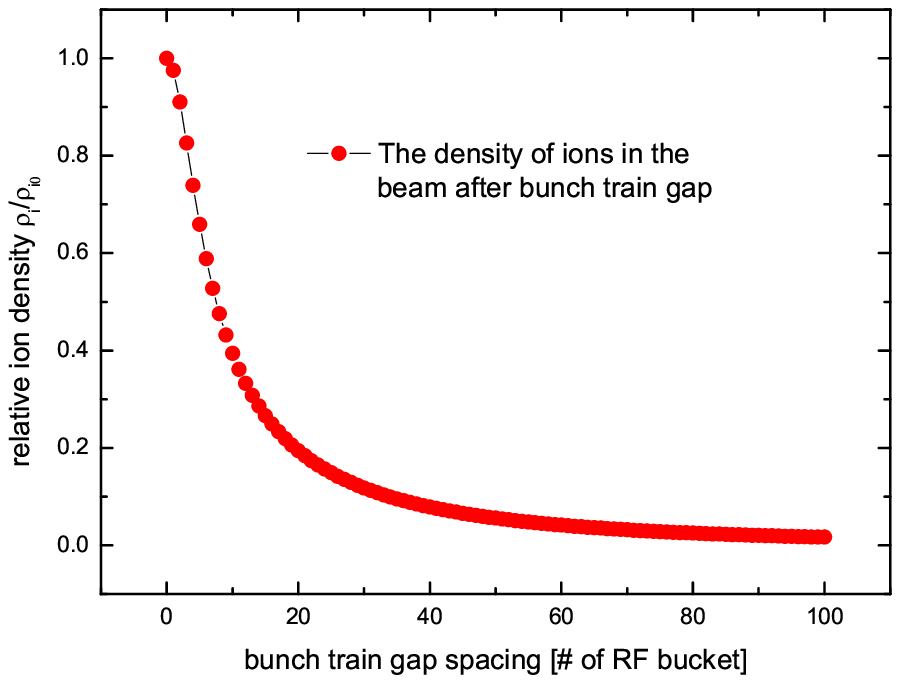}}
\caption{The density of the residual ions in the beam after the
bunch train gap.} \label{Figure:3}
\end{minipage}
\hspace*{0.2cm}
\begin{minipage}{0.5\columnwidth}
\centerline{\includegraphics[width= 0.9\columnwidth ]{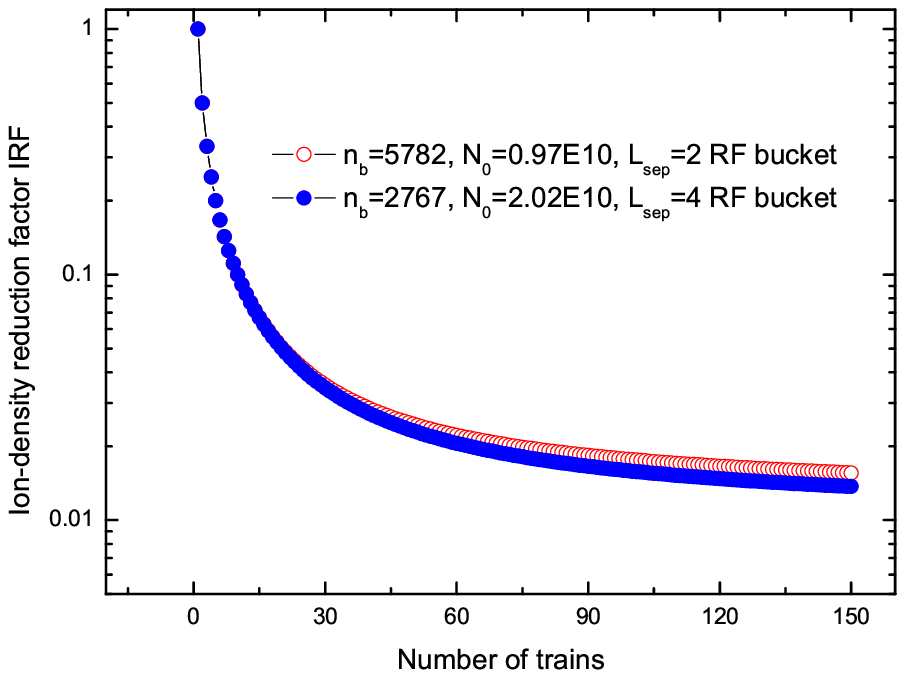}}
\caption{IRF factor in OCS6 damping ring for fill pattern \textsl{A}
and \textsl{C}.} \label{Figure:4}
\end{minipage}
\end{figure}
where $N_{train}$ is the number of trains, $\tau_{ion}$ is the
diffusion time of the ion cloud which can be estimate from Eq.(3).
\textsl{IRF} is the ratio of the ion density with gaps and without
gaps. In one long bunch train case, the ring is completely filled
and the ions can accumulate indefinitely. With a fixed gap, a larger
number of shorter bunch train helps to keep the ion density low.
However, for a fixed ring circumference and total number of bunches,
the length of gap shrinks as the number of bunch trains increases.
The optimum fill pattern depends on the diffusion time, the
circumference, and number of bunches. Figure 4 shows the
\textsl{IRF} versus number of trains in OCS6 damping ring for fill
pattern case \textsl{A} and \textsl{C} respectively. It can be seen
here if the harmonic number and ring circumference are fixed, the
\textsl{IRF} reduces with respect to the number of trains. Beyond 60
trains the \textsl{IRF} does not change a lot.

The beam parameters of the ILC damping ring are listed in Table
2.4-1 of Ref. \cite{Aarons}. There are 5 different fills in this
ring. For different fill patterns, the total number of particles is
kept constant so that the specified luminosity can be achieved.

\begin{figure}[htpb]
\begin{minipage}{0.5\columnwidth}
\centerline{\includegraphics[width= 0.9\columnwidth ]{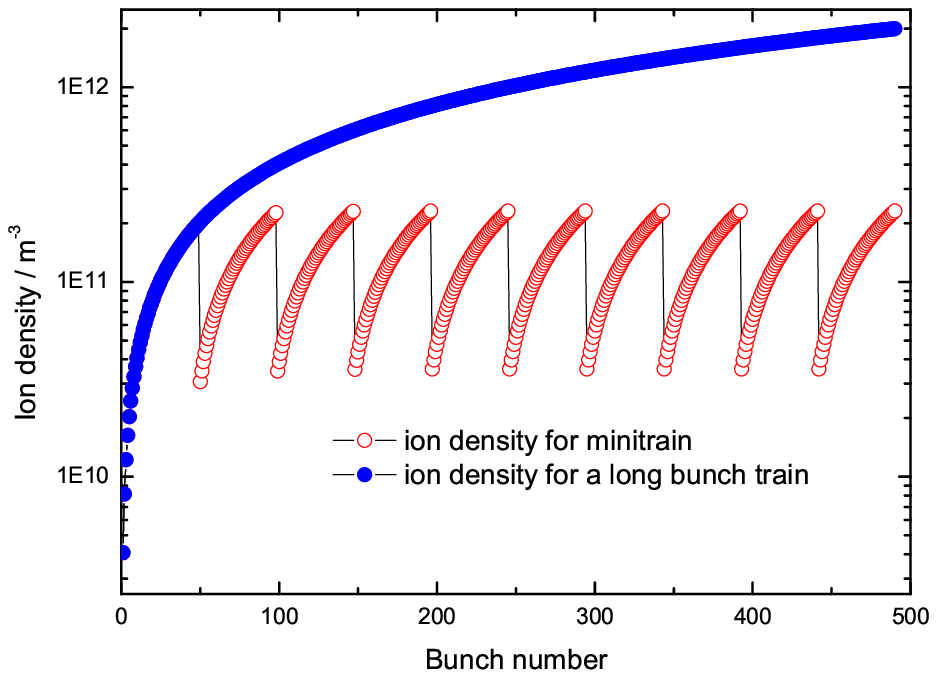}}
\caption{Ion density in fill pattern case \textsl{A}.}
\label{Figure:5}
\end{minipage}
\hspace*{0.2cm}
\begin{minipage}{0.5\columnwidth}
\centerline{\includegraphics[width= 0.9\columnwidth ]{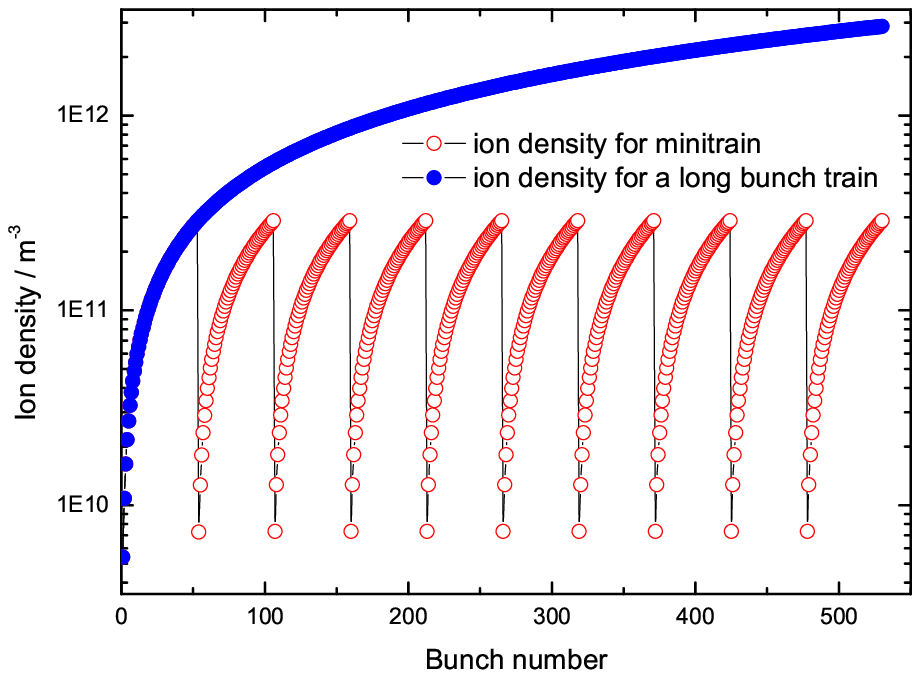}}
\caption{Ion density in fill pattern case \textsl{B}.}
\label{Figure:6}
\end{minipage}
\end{figure}
\begin{wrapfigure}{r}{0.5\columnwidth}
\centerline{\includegraphics[width=0.45\columnwidth]{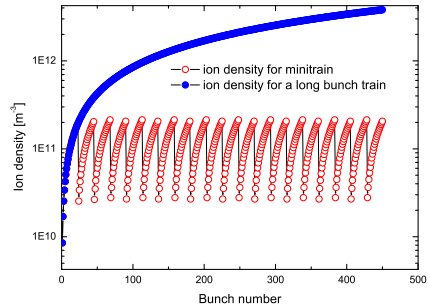}}
\caption{Ion density in fill pattern case
\textsl{C}.}\label{Figure:7}
\end{wrapfigure}

The ion density for different fill patterns case \textsl{A},
\textsl{B} and \textsl{C} are shown in Figure 5, Figure 6 and Figure
7 respectively for ten bunch trains. The CO partial pressure is 1.0
nTorr. It can be seen that the ion density for a single long bunch
train will increase linearly with respect to the bunch number.
However, if the gaps are introduced between the adjacent bunch
trains, the ion density is reduced significantly. It also indicates
that the ion density for mini-trains can quickly reach the peak
value after the first few bunch trains. For the fill patterns
\textsl{A}, \textsl{B} and \textsl{C}, the mini-trains can reduce
the ion density by about two orders of magnitudes comparing to a
single long bunch train. Since the growth rate of FII is
proportional to the ion density, it indicates that the growth rate
of FII can be reduced by a factor of 100. In this case, the FII can
be potentially damped by a fast feedback system.

\section{Simulation study of FII in the ILC damping ring}
  A weak-strong code is used to simulate the FII in the ILC electron
damping ring \cite{Xia06}. The effect of mini-trains is taken into
account. Figure 8 shows the growth of the vertical oscillation
amplitude versus number of turns for a single long bunch train and
for mini-trains of fill pattern \textsl{A}. The $5782^{nd}$ bunch is
recorded here.
% insert table
\begin{figure}[htpb]
\begin{minipage}{0.5\columnwidth}
\centerline{\includegraphics[width= 0.9\columnwidth ]{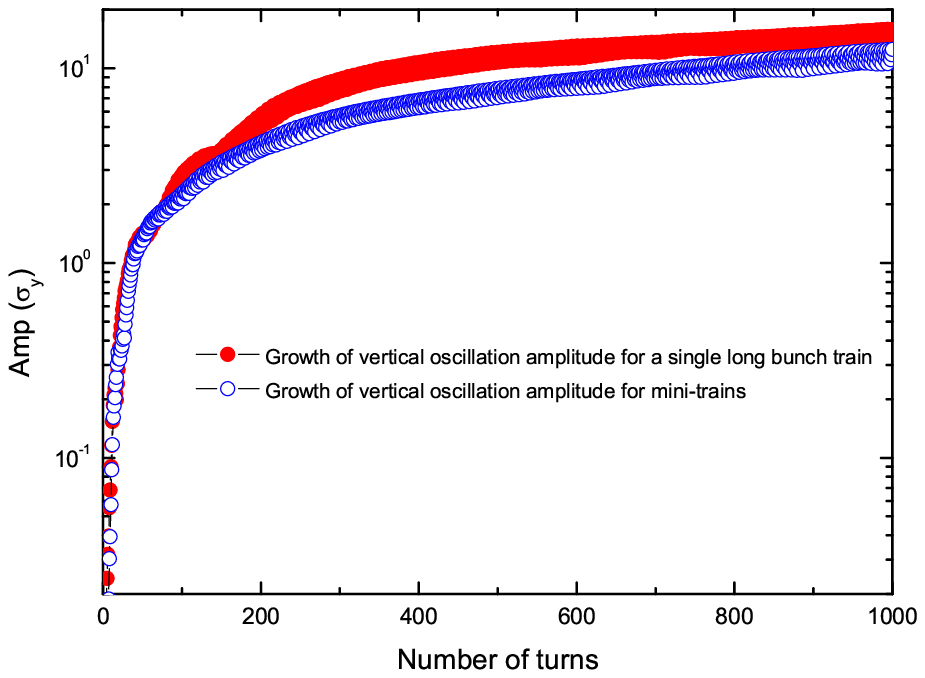}}
\caption{Growth of vertical oscillation amplitude for a single bunch
train and mini-trains in fill pattern case \textsl{A} in a CO
pressure of 1nTorr.} \label{Figure:8}
\end{minipage}
\hspace*{0.2cm}
\begin{minipage}{0.5\columnwidth}
\centerline{\includegraphics[width= 0.9\columnwidth ]{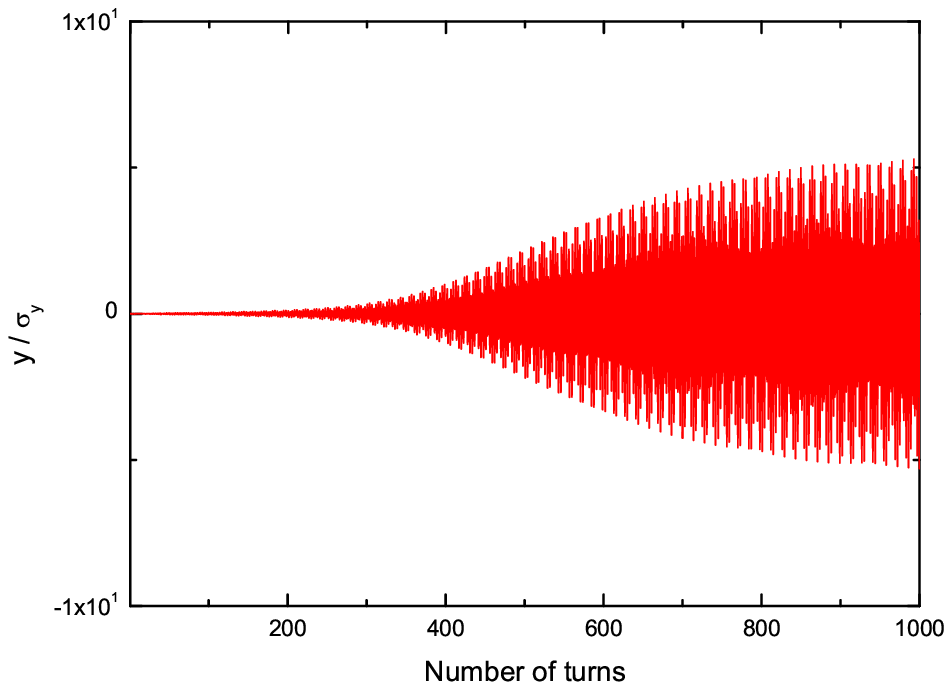}}
\caption{The $5782^{nd}$ bunch centroid oscillation in the fill
pattern case \textsl{A}.} \label{Figure:9}
\end{minipage}
\end{figure}
% end of table insertion
It can be seen when the gap is introduced in the bunch trains, the
growth of vertical oscillation slows down. This is because in the
case the ion density for mini-trains is less than that of a single
long bunch train case in Figure 5. Figure 9 shows the bunch centroid
oscillation versus number of turns in the fill pattern case
\textsl{A}. The $5782^{nd}$ bunch is recorded here. We can see the
bunch centroid begins to oscillate with small amplitude and then
reaches the saturation after about 600 turns \cite{Xia07}. This is
also one of the characteristics of the FII.
\section{Conclusion}
  Gaps between bunch trains can significantly reduce the ion density
in the ILC damping ring; a gap exceeding 30 RF bucket reduces the
ion density by a factor 10. Depending on fill pattern the ion
density diminishes by about two orders of magnitude compared to one
long bunch train case. Simulation shows the growth of vertical
oscillation amplitude to be attenuated with gaps in the fill.

% ****************************************************************************
% BIBLIOGRAPHY AREA
% ****************************************************************************
\begin{footnotesize}
% IF YOU DO NOT USE BIBTEX, USE THE FOLLOWING SAMPLE SCHEME FOR THE REFERENCES
% ----------------------------------------------------------------------------

% ----------------------------------------------------------------------------
% ****************************************************************************
% END OF BIBLIOGRAPHY AREA
% ****************************************************************************
\end{footnotesize}

\begin{thebibliography}{99}
% Please replace the numbers for   contribId   and   sessionId
% in the following URL. You can get this information by going to
% http://indico.cern.ch/confAuthorIndex.py?confId=9499
% and search for your contribution and click on the title
% Be aware: '&amp;' must be replaced by simple '&' as in example below
\bibitem{url} Slides: \\
\verb$http://ilcagenda.linearcollider.org/contributionDisplay.py?contribId=334&sessionId=65&confId=1296$
%------- replace following references ;-)
\bibitem{Byrd} J.M.~Byrd {\it et~al.}, SLAC-PUB-7389 (1996).
\bibitem{Tor}  T.~Raubenheimer and F.~Zimmermann, Phys. Rev. E {\bf 5} 5487 (1995).
\bibitem{Stupakov} G.V.~Stupakov, T.~Raubenheimer and F.~Zimmermann, Phys. Rev. E {\bf 5} 5499 (1995).
\bibitem{Wolski} A.~Wolski {\it et~al.}, LBNL-59449 (2006).
\bibitem{Kuriki} M.~Kuriki, {\it et~al.}, ILC-Asia-2006-03 (2006).
\bibitem{Lanfa}  L.~Wang, talks at ILCDR07 Workshop in Frascati, (2007).
\bibitem{Aarons} G.~Aarons, {\it et~al.}, ILC-Report-2007-01 (2007).
\bibitem{Wang}   L.~Wang, {\it et~al.}, SLAC-PUB-12643 (2007).
\bibitem{Xia06}  G.~Xia, {\it et~al.}, Proceedings of EPAC06, MOPLS133, Edinburgh, UK (2006).
\bibitem{Xia07}  G.~Xia, {\it et~al.}, Proceedings of PAC07, THPMN016, Albuquerque, New Mexico, USA (2007).
\end{thebibliography}
\end{document}